\documentclass[onecolumn]{aastex63}

\shorttitle{Cepheid Metallicity: Theory} \shortauthors{Madore, Freedman \& Owens}

\begin{document}

\title{Chicago-Carnegie Hubble Program (CCHP) \\A Multi-Wavelength Search for the Effects of Metallicity on the Cepheid Distance Scale.\\ Part II: Theoretical Models and Synthetic Spectra}

\correspondingauthor{Barry F. Madore}

\medskip
\author[0000-0002-1576-1676]{\bf Barry F. Madore} 
\email{barry.f.madore@gmail.com}
\affil{The Observatories Carnegie Institution
for Science 813 Santa Barbara St., Pasadena, CA ~~91101 }
\affil{Department of Astronomy \& Astrophysics, University of Chicago, 5640 South Ellis Ave., Chicago, IL 60637}

\medskip
\author[0000-0003-3431-9135]{\bf Wendy L. Freedman} 
\email{wfreedman@uchicago.edu}
\affil{Department of Astronomy \& Astrophysics, University of Chicago, 5640 South Ellis Ave., Chicago, IL 60637}\affiliation{Kavli Institute for Cosmological Physics, University of Chicago,  5640 South Ellis Ave., Chicago, IL 60637}
\author[0000-0003-3339-8820]{\bf Kayla Owens} 
\email{kaowens@uchicago.edu}
\affil{Department of Astronomy \& Astrophysics, University of Chicago, 5640 South Ellis Ave., Chicago, IL 60637}\affiliation{Kavli Institute for Cosmological Physics, University of Chicago,  5640 South Ellis Ave., Chicago, IL 60637}

\begin{abstract}

This is the second of two papers exploring the effects of metallicity on the multi-wavelength properties of Cepheids in terms of their multi-wavelength period-luminosity (PL) relations, impacting their use as extragalactic distance indicators, underpinning one of the most popular paths to estimating of the expansion rate of the Universe, $H_o$. 

In Paper I (Madore \& Freedman 2024) we presented five {\bf tests} for the influence of metallicity on galactic and extragalactic Cepheid PL relations, spanning nearly 2~dex in metallicity, and inspecting PL relations from the optical (BVI), through the near-infrared (JHK) and into mid-infrared (at 3.4 and 4.5 microns). And,in no case were any statistically significant results forthcoming. Here we interrogate published spectral energy distributions constructed from theoretical (static) stellar atmospheres, covering the surface gravity and temperature ranges attributed to classical (supergiant, F and K spectral type) Cepheid variables, and explore the differential effects of changing the atmospheric metallicity, down by 2 dex from solar (a factor of 100 below the average Milky Way value) and then up from solar by 0.5 dex (i.e., factor of 3x above the Milky Way value). 

The theoretical models clearly show that metallicity systematically impacts each of the bandpasses differentially: the level of this effect is largest in the ultraviolet (where line blanketing is most intense), reversing sign in the optical (due to flux redistribution from the UV), and then asymptotically falling back to zero from the red to the far infrared. The discovered effects of metallicity are systematic, but they are small; and as such they do not contradict the findings of Paper I, but they do explain why the problem has been so hard to resolve given the low level of precision of the photometry for all but the very nearest and apparently brightest Cepheids. 

The most interesting (and useful) result of this investigation is that from a close examination of the models we have discovered infrared regions in Cepheid spectral energy distributions that allow for the simultaneous correction for extinction and metallicity differences by the judicious combination of only two filters: one in the near-infrared and one in the mid-infrared, the former centered at 1.2 microns and the other centered at 3.6 microns. Even more exciting and promising is the fact that both of these bands can be simultaneously observed in single exposures using the dual-channel imager, NIRCam on JWST.

\end{abstract}

 .


\section{Introduction}

In the first paper in this investigation (Paper I, Madore \& Freedman 2024) we undertook a series of tests searching for evidence for a metallicity term in any of the multi-wavelength Period-Luminosity (PL) relations currently available for Cepheids in Milky Way and in neighboring galaxies, close enough to have high-precision data available.  
Five tests were undertaken. The first used trigonometric parallaxes to individual Galactic Cepheids obtained by Gaia and/or HST. Inconsistencies in the two data sets were resolved by requiring that the run of the observed widths of the multi-wavelength PL relations conform numerically to the known widths which have been independently calibrated to be monotonically declining functions of increasing wavelength. This (moderately high-metallicity) Milky Way calibration was then used to predict the distance to the measurably-lower metallicity sample of Cepheids in the Large Magellanic Cloud (LMC). That distance was found to agree with the detached eclipsing binary star (geometric) distance at the millimag level. This first result left no statistically significant room for a metallicity correction. The second test examined the wavelength-dependent residuals obtained by subtracting the observed apparent distance moduli from the globally fitted extinction curve. If metallicity effects are wavelength dependent, their signature should be imprinted on their residuals from the extinction-curve fit. None were found in the highest precision datasets, nor in samples more restricted in wavelength coverage and/or in data with lower precision apparent moduli. A reprise of a long-standing test for a metallicity term in the 
optical Wesenheit (V,V-I) reddening-free distance moduli (Freedman \& Madore 2022) also shows no statistically significant correlation of the difference between the Cepheid and TRGB distances to over two dozen galaxies covering a 2.0~dex range of metallicities centered approximately on the Milky Way metallicity.
That the larger sample of Milky Way field Cepheids (that have published photometry and Gaia parallaxes) also have a wide (1 dex) range of published metallicities, allows for yet another test looking for trends with metallicity of the magnitude deviations from the mean PL relation. This test was first undertaken by Narloch et al. (2023) using Sloan PL relations, and it is repeated here in this paper using more conventional (Kron-Cousins, and Johnson) filter sets. The finding are identical: large scatter and no detectable trend with metallicity at any of the observed bands. 

These tests are largely independent. But all of them came to the same conclusion: {\it To the limit of the precision of the existing data, no statistically significant metallicity term can be found at any wavelength }. . . with one outstanding exception. In the longest bandpass tested, the 4.5 micron filter (available in the now terminated Spitzer mission) there is a CO band head that crosses the whole [4.5] filter. CO is both a strong absorber of the continuum and its relative strength at any given temperature is very sensitive to metallicity. Its existence proves that the observed Cepheids do have different metallicities and that these difference can be (and have been) detected in their individual spectra. The CO features are statistically very significant. But, no other features searched for to date come close.  

\section{Going Beyond the Observations}
The null results of Paper I, as reviewed above, are not the end of the story. In this paper, we turn to theory for an independent perspective on the impact of metallicity variations on the emerging flux passing through the atmospheres of Cepheid variables. Simply put, if a broadband measurement is made of a region in the Cepheid's spectrum where there is an abundance of metallic-line transitions, there will be suppression of the continuum, making the star {\it fainter} at that magnitude as a function of increasing metallicity, {\it ceteris paribus}, in this case, at fixed temperature and surface gravity. The photons removed by metallic ``line blanketing'' will eventually be thermalized and rejoin the ambient photon field, redistributed over the entire spectral energy distribution (SED) at the characteristic temperature of the star. However, if significant amounts of energy are removed in the ultraviolet, say, then {\it measurable increases} in the continuum will be centered at longer wavelengths (around 6000A, slightly long-ward of the V band, in the case of Cepheids). Thus, the impact of increasing metallicity, as seen in the blue and ultraviolet, ultimately {\bf results} in making Cepheids brighter in the visible and red, with declining contributions to the near- and far-infrared.
However, not all absorption is due to atomic species; molecules forming in these same atmospheres will have lower-energy (rotational and vibrational) transitions, which will be predominantly found in the near and far infrared. For Cepheids CO is known to have very wide (1000A) band heads centered at 4.5 and 2.2 microns. The energy removed by those bands will also be thermalized, but redistributed to bluer wavelengths, again contributing to making Cepheids brighter around the V band as metallicity increases, with the stars becoming fainter in specifically impacted regions in the near- and mid-infrared. 

The above outline is demonstrated in detail with the construction of stellar atmospheres, treating thousands of lines and bands resulting from controlled amounts of metals, drawn from across the periodic table according to known relative cosmic abundances. For our purposes we have chosen the PHOENIX model atmospheres (Husser et al. 2017) to quantify the effects of varying the total abundance of metals on the spectral energy distributions of stars, with models having effective temperatures and surface gravities characteristic of Cepheids: i.e., intermediate temperatures of T = 5000 to 6000K, and low surface gravities in the range of log g = 3.0 to 3.5.  

We start with Figure 1. On the vertical axis is the difference (expressed in magnitudes) between a high-metallicity SED and a low metallicity SED for a star having a temperature of 5000K  and a surface gravity of log g = 3.0, where the metallicity difference is 1.5 dex. For reasons that will become clear shortly, the horizontal axis is given in units of inverse microns. The locations of broadband filters, UBVI in the optical, JHK in the near-infrared and 3.6 and 4.6 microns in the mid-infrared are highlighted by thin, black, vertical lines, annotated accordingly. Dashed blued and black lines will be explained later in the text. The thin black horizontal line, crossing the figure at $\Delta$~mag = 0.0, is in fact the trace of the spectrum of a star that has no metals in its atmosphere. Portions of the red curve falling above that line have excess luminosity, (due to thermalized flux redistribution) compared to a zero-metallicity star. Portions falling below that line suffer from metallic line and molecular band absorption, in the blue/ultraviolet (to the right, beyond V) and in the NIR/MIR  (to the left, beyond J), respectively.
Circled points highlight the metallicity-induced deflections that are recorded for the various broadband filters commonly used in the construction of Cepheid PL relations, the only exception being the U band which is not generally considered for distance scale work, for a variety of practical reasons. 
\begin{figure*}[hbt!] 
\centering
\includegraphics[width=8.5cm, angle=-0] {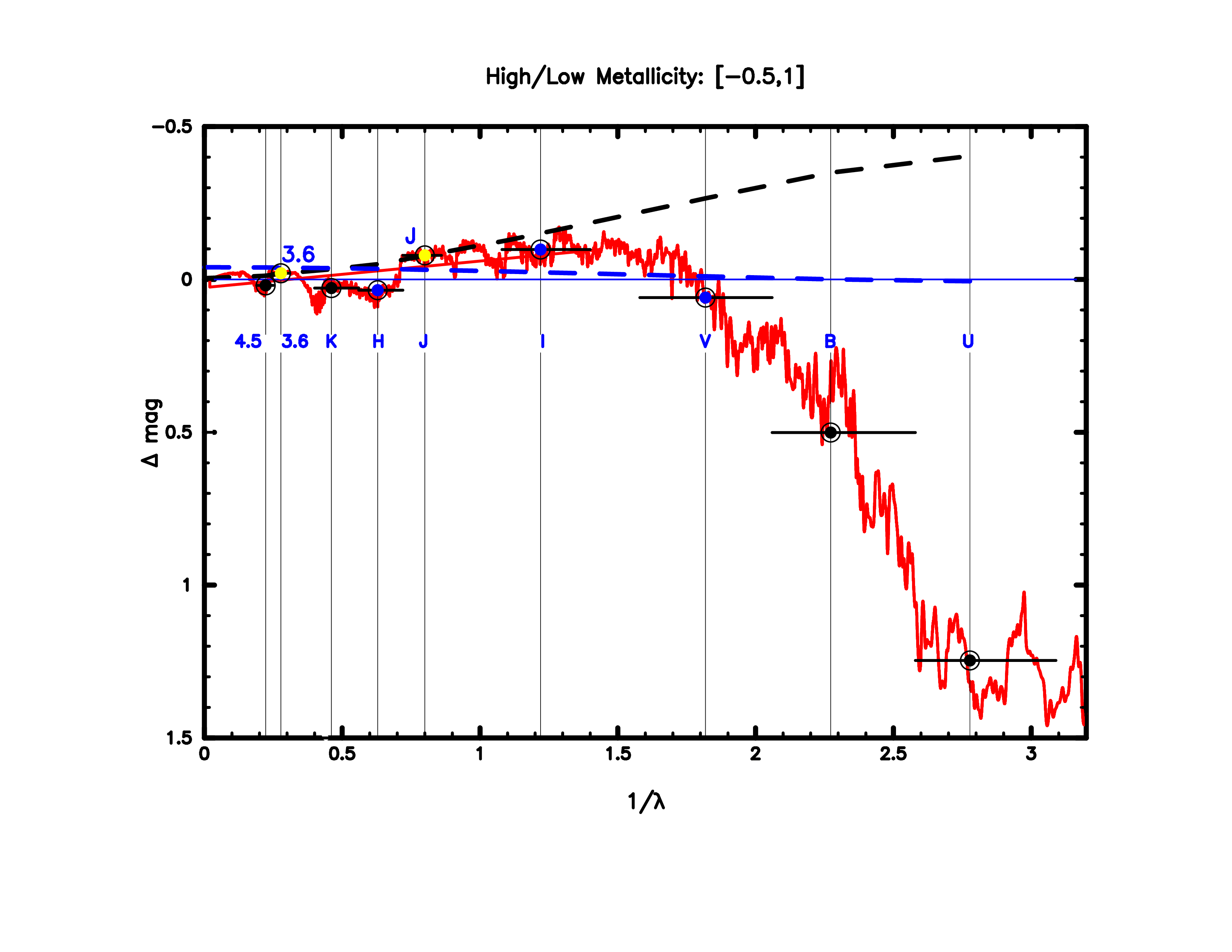}
\includegraphics[width=8.5cm, angle=-0] {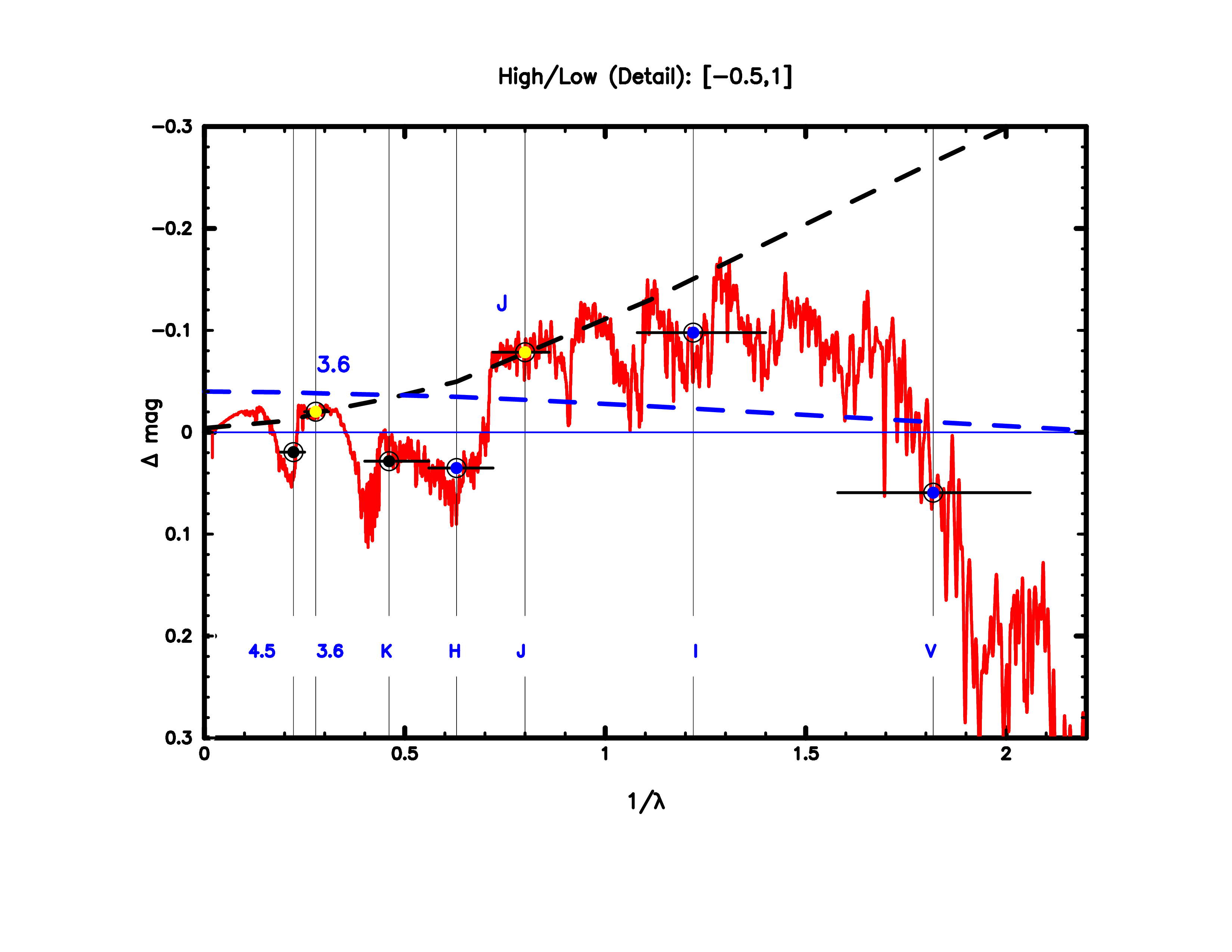}
\caption{\small -- Left Panel: Magnitude differences of two G0 supergiant spectra having metallicites of -0.5 and 1.0 respectively, as a function of inverse wavelength. Filled circles represent broadband magnitudes integrated over the wavelength intervals indicated by the horizontal black lines. J and 3.6 micron data points are emphasized in yellow. The curved broken black line having the form of a Cardelli extinction curve passing exactly through those two data points and extrapolating to zero offset at infinite wavelength. The magnitudes and signs of these selected points define the zero-point offsets that need to applied when correcting any give PL relation for atmospheric metallicity effects. They cross zero several times,and they range from more than +1.0~mag in the U band and +0.5 ~mag in the B band, to -0.1~mag for filters redward of the V band to the J band, dropping back below the line with finite-width structures at 1.6 (H), 1.2 (K) and 4.5~microns. The right panel shows a slightly zoomed-in view of the same magnitude-differenced pair of SEDs, showing more clearly the two CO bandheads: one at 4.5~microns and the other near to the K band at 2.2 microns}
\end{figure*}

\subsection{First Indications of the Predicted Metallicity Effects}

We now return to the plot of apparent moduli as a function of inverse wavelength for the LMC Cepheids given in Figure 1 of Paper I. 
The data are replotted here in the top portion of Figure 2. The thin red line is the E(B-V) = 0.105~mag  fit of a Cardelli extinction curve to the seven apparent distance moduli (excluding the point at 4.5 microns).
The yellow data points, in the lower subpanel are the result of subtracting the observed apparent moduli (in the upper panel) from an adjusted extinction curve (shown by the broken black line) moving E(B-V) from 0.105~mag to the higher value of  0.118~mag. 
The exact adjustment in reddening was driven by fitting the extinction curve to just two bands: J and [3.6]. 
This results in a simultaneous fit to both the true line-of-sight extinction and the revelation of the metallicity-induced residuals caused by the lower metallicity of the Large Magellanic Cloud (LMC) Cepheids compared to the MW calibration. 
The thick black line in the sub-panel below this plot shows an exemplary ratioed synthetic spectrum that tracks the yellow residuals.  
As noted in the upper left corner of the main plot, the doubly-corrected true distance modulus to the LMC found here is 18.485~mag, which agrees exceedingly well with the detached eclipsing binary (DEB) distance modulus (18.477~mag), at the millimag difference level of only 0.008~mag.

We now justify this surprising turn of events and the newly suggested methodology by turning to a broader view of the metallicity plot first given in Figure 1, in this case examining it over a wider range of metallicities and then sampling the full temperature range defining the Cepheid instability strip (5000 to 6000K).

The upper right panel in Figure 3 shows the synthetic PHOENIX spectrum for a supergiant star, having a surface temperature of 5000K and an extremely low metallicity of [Fe/H] = -2.0 dex.
That spectrum divided by its high-metallicity ([Fe/H] = +1.0~dex) analog is shown in the lower left panel, where differences are expressed in units of magnitudes.
This plot is for the highest contrast of metallicities, as is readily attested to be the depth of the CO band head at 4.5~microns to the left of the vertical blue line indicating the position of the 3.6~micron filter sampling the continuum in this and each of the other panels. 
The other blue line, further to the right, marks the position of the J-band filter, which cleanly samples continuum flux. Both of these infrared filters are marked by circled cyan points. 
Moving up in the figure, and from right to left we have three more spectral-ratio plots, each normalized to [Fe/H] = +1.0~dex and progressively decreasing in the comparison metallicity going from [Fe/H] = -0.5~dex (LMC-like metallicity,  middle-left sub-panel), to [Fe/H] = 0.0~dex (solar metallicity,  middle-right sub-panel), to [Fe/H] = +0.5~dex (top-left sub-panel).

The two curving red lines in each of the spectra in Figure 3 are reddening curves, scaled and fit to the VI data alone and then to the VIH data. The former is essentially the Wesenheit (V,V-I) solution for the reddening-corrected true distance modulus (found at the curve's extrapolation to $1 / \lambda$ = 0.00). The latter is a hybrid Wesenheit (H,V-I) solution. It has to be emphasized here that each of these means of combining selected filters, in order to correct for reddening, are also affected by metallicity. And, the full range of the effect is shown by the error bars to the far left of each of the panels. The range due to metallicity is 0.048~mag for W(V,V-I) and 0.035~mag for W(H,V-I). Both effects are small but systematic, trending toward brighter magnitudes with increasing metallicity. 

\section{Introducing W(J,36)}

The cyan curves in each of the plots are also extinction curve fits, but this time to a novel pair of bands, J at 1.2~microns and M at 4.5~microns, that have a unique, combined response simultaneously to reddening and metallicity. 
That is, these two bands change in lock step with changes in metallicity and/or extinction. In forming W(J,36) = J - R$_J~ \times $ (J - 3.6),  one designs this Wesenheit function to correct for one effect (reddening, say) and it automatically corrects for the other (metallicity), in this case. Moreover, there is no bias/offset in the intercept of this functional extrapolation to $1 / \lambda$ = 0.00), where the true distance modulus is being recovered. 
For any given Cepheid, changing its reddening and/or its metallicity results in no change in its W$_{J36}$ magnitude. This is graphically illustrated in each of the plots in Figure 3  (and indeed, in Figure 2) where the long-wavelength intercepts of the W(J,36) cyan lines in each of the plots are all identically zero, despite the fact that the metallicity changes by 2.5~dex! 

In Figure 4 we illustrate the remarkable stability of the pairing (J,3.6) with changes in the metallicities by color-coding and overlaying the four plots in Figure 3 to more easily visualize the systematics of metallicity change at fixed temperature and surface gravity.
The bottom panel shows lowest metallicity spectrum in yellow and the highest metallicity spectrum in black. The trends are clear and the effect is especially well seen in the two metallicity sensitive CO bands at 4.5 and 1.2 microns.
A closer inspection suggests to the eye that these spectra may be simply scaled version of each other.

The upper panel in Figure 4 proves that that impression is largely true. By scaling the amplitudes of the 4.5~micron CO feature to the highest metallicity (black) spectrum, and shifting them vertically into registration, results in the composite spectrum in the upper panel. Three aspects of this plot are important to note: (1) From 1/$\lambda$ = 0.0 to 0.45, and again from 1/$\lambda$ = 0.7 to 2.5, the CO scaling returns a universal (minimum scatter) spectrum. (2) The region from 1/$\lambda$ = 0.45 to 0.7 defies the scaling and actually inverts its response to metallicity, compared to its dispersion seen in the lower panel. (3) The J and 4.5 micron continuum points scale along with the CO features and preserve their unique attribute of being fit by a extinction curve passing through zero at infinite wavelength.

\section{Returning to Data }

\subsection{Deriving and Testing the Absolute Calibration of W(J,3.6)}
At this point in time there are only a few Cepheid data sets that include both near-infrared, J band and mid-infrared 3.6~micron light curves and mean magnitudes. The best of those are a Milky Way sample, the LMC and IC~1613. The first two galaxies have geometric zero points, and the third galaxy represents two extremes: being of very low metallicity and having very low total line-of-sight extinction. We now consider each of these systems in turn.

\subsubsection{W(J,36) and the Milky Way Cepheid Zero-Point Calibration}

We proceed here to use the Milky Way Period-Luminosity relations derived in Paper I to calibrate zero point and scatter of the W(J,36) Wesenheit function, fully defined here for the first time:

$$ W(J,36) = J - R_J \times (J - [3.6]) $$ 

where R$_J$ = A$_J$/E(J - [3.6]) = 0.242.

Figure 5 shows the J and [3.6] micron PL relations for the 17 Milky Way Cepheids having measurement in both bands, color-coded by the two independent parallax samples Gaia, yellow and HST, blue). Above them (displaced for better visibility) is the W(J,36) PL relation. It has a scatter of 0.20~mag and a zero point of -5.58~mag measured at log P = 1.0.
The statistical error on that zero point is $\pm$0.20/$\sqrt{17}$ = 0.049~mag. 

\subsubsection{W(J,36) and the LMC Cepheid Zero-Point Calibration}

Figure 6 (upper panel) shows the J and [3.6] micron PL relations for the larger sample of 88 Cepheids in the LMC having mean magnitudes measured in both bands. Above them (again, displaced for better visibility) is the W(J,36) PL relation. It has a scatter of 0.20~mag and a zero point of -5.55~mag measured at log P = 1.0.
The statistical error on that zero point is $\pm$0.20/$\sqrt{82}$ = 0.022~mag. The DEB distance to the LMC was used to establish the absolute magnitude of this W (J,36) calibration.
\subsubsection{W(J,36) and the IC~1613 Cepheid Zero-Point Calibration}

Figure 6 (lower panel) shows near- and mid-infrared PL relations for IC~1613 Cepheids derived from ground-based photometry in J (red points) combined with Spitzer 3.6 micron data (blue points) in the lower portion of the figure. The W(J,3.6) Wesenheit function is shown in the middle of the plot: black points, fit by a thin solid black line flanked by $\pm$~2-sigma dashed lines. The yellow points at the top of the plot trace the run of (J-3.6) color with log P, but mainly serve to show the decreased precision in overall photometry in the IC~1613 sample as compared to the equivalent data plotted in the panel to the left for the LMC data.

\subsubsection{W(J,3.6): Combining the Zero Points}

Weighting the two independently derived zero points by the inverse square of their statistical errors we arrive at a final calibration of
$$W(J,3.6) = -3.19 ~(log P - 1.0) - 0.242 \times (J - [3.6]) -5.45 ~\pm 0.031 ~(se)$$
\clearpage
\subsection{First Test of the W(J,36) Method}

\begin{figure*}[hbt!] 
\centering
\includegraphics[width=18.5cm, angle=-0] {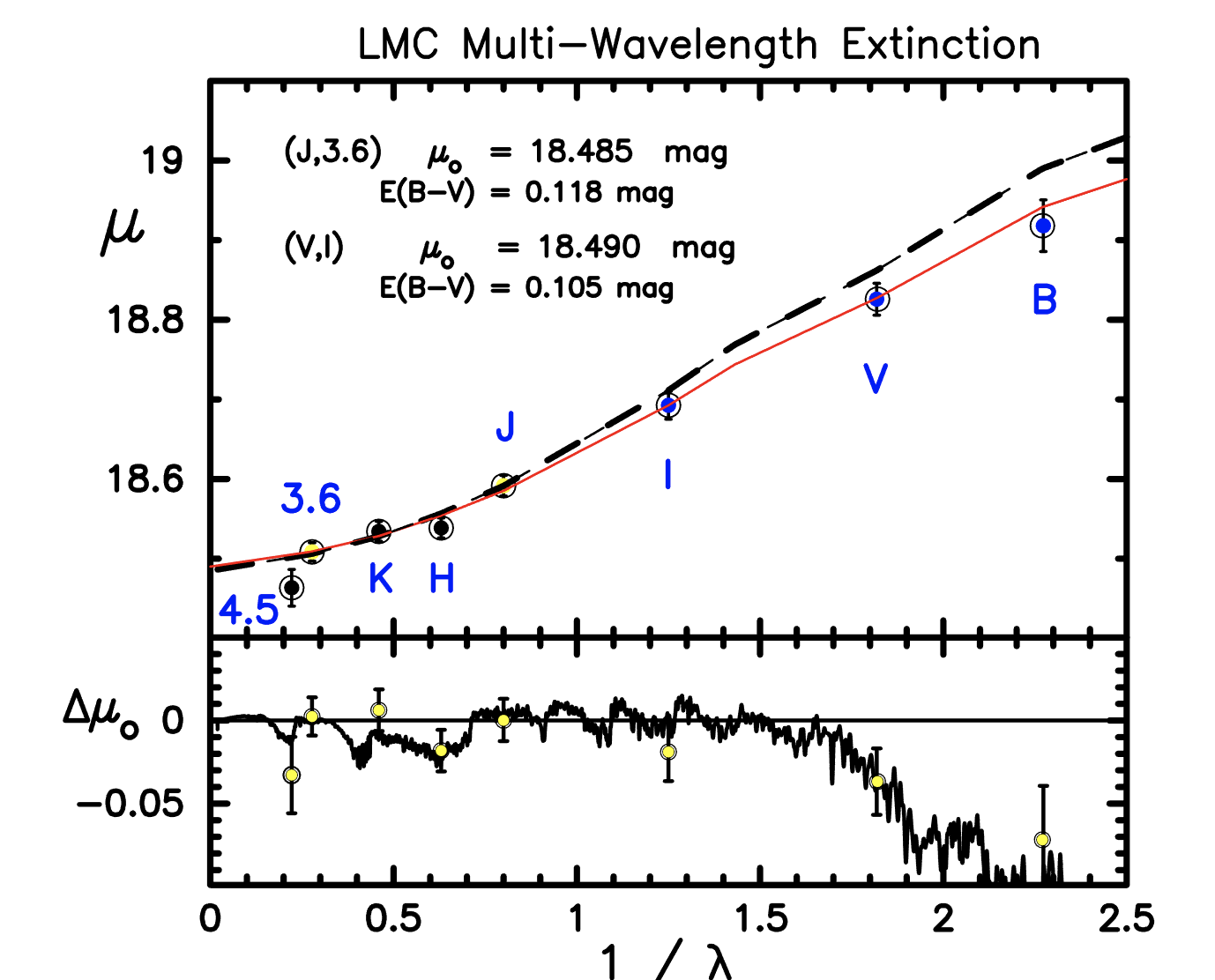}
\caption{\small -- Multi-Wavelength Extinction Plot and Residuals for LMC Cepheids. In the upper plot, the thin red line is the E(B-V) = 0.105~mag fit of a Cardelli extinction curve to the seven apparent distance moduli (excluding the point at 4.5 microns). 
The broken black line is the extrapolated, E(B-V) = 0.118~mag fit to the two yellow circled data points at 3.6 and 1.2 (J-band) microns.
The yellow data points with error bars, in the lower subpanel, are the result of subtracting the observed apparent moduli (in the upper panel) from the adjusted (steeper) extinction curve. 
The exact adjustment in reddening was driven by fitting the extinction curve to just two bands: J and [3.6]. 
This procedure results in a new fit to the true line-of-sight extinction and the revelation of the metallicity-induced residuals caused by the lower metallicity of the LMC Cepheids compared to the MW calibration. 
The thick black line in the subpanel below this plot shows an exemplary ratioed synthetic spectrum that now tracks the yellow residuals.  
Corrected for reddening and metallicity, the true distance modulus to the LMC found here is 18.485~mag, which at the millimag level agrees the DEB distance modulus of 18.477~mag (Pietryznski et al. 2019).}
\end{figure*}

\begin{figure*}[hbt!] 
\centering
\includegraphics[width=18.5cm, angle=-0] {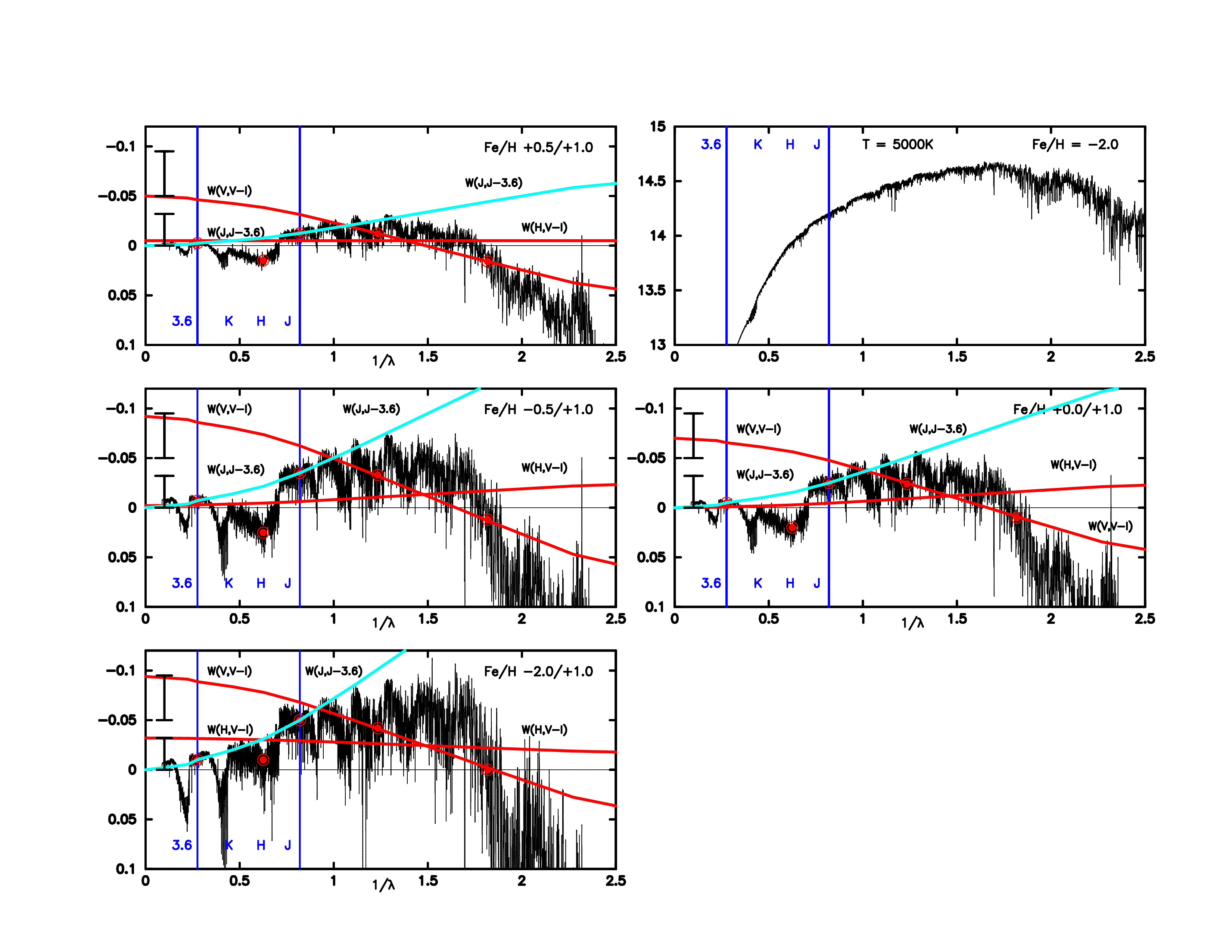}
\caption{\small -- Optical to infrared model spectra of a 5,000K supergiant with a selection of atmospheric metallicities ranging [Fe/H] = -2.0~dex (lower-left panel) to [Fe/H] = +0.5~dex (upper-left panel), each normalized to the highest metallicity spectrum having [Fe/H] = +1.0~dex. Upper-right panel shows the lowest-metallicity spectrum for the un-normalized star having [Fe/H] = -2.0 dex, least influenced by atomic or molecular features. All spectra are plotted as a function of inverse wavelength. The vertical blue lines mark the central wavelengths of broadband J and 3.6 micron filters. (a) The upward-trending cyan curve is an interstellar extinction curve scaled and fit only to the J and 3.6 micron points in each of the spectra. All of those fits extrapolate to zero offset at infinite wavelength, leading to no dependence of that modulus to both reddening and metallicity. (b)The downward trending red line is an interstellar extinction curve, independently scaled and fit to just the V and I  data points. Their intercepts systematically extrapolate to non-zero offsets that range from -0.050 to -0.095~dex, leading to a reddening-corrected modulus whose zero point has a metallicity slope of $\mu$ = -0.045/2.5 = -0.018~mag/dex. (c) The relative flat red curve in the middle of each plot is a fit to the VIH data. Their intercepts are also systematically offset from zero, but have a slightly smaller range than the VI solutions. That range is -0.000 to -0.032 mag giving a metallicity slope of $\mu$ = -0.032/2.5 = -0.013~mag/dex.}
\end{figure*}

\begin{figure*}[hbt!] 
\centering
\includegraphics[width=18.5cm, angle=-0] {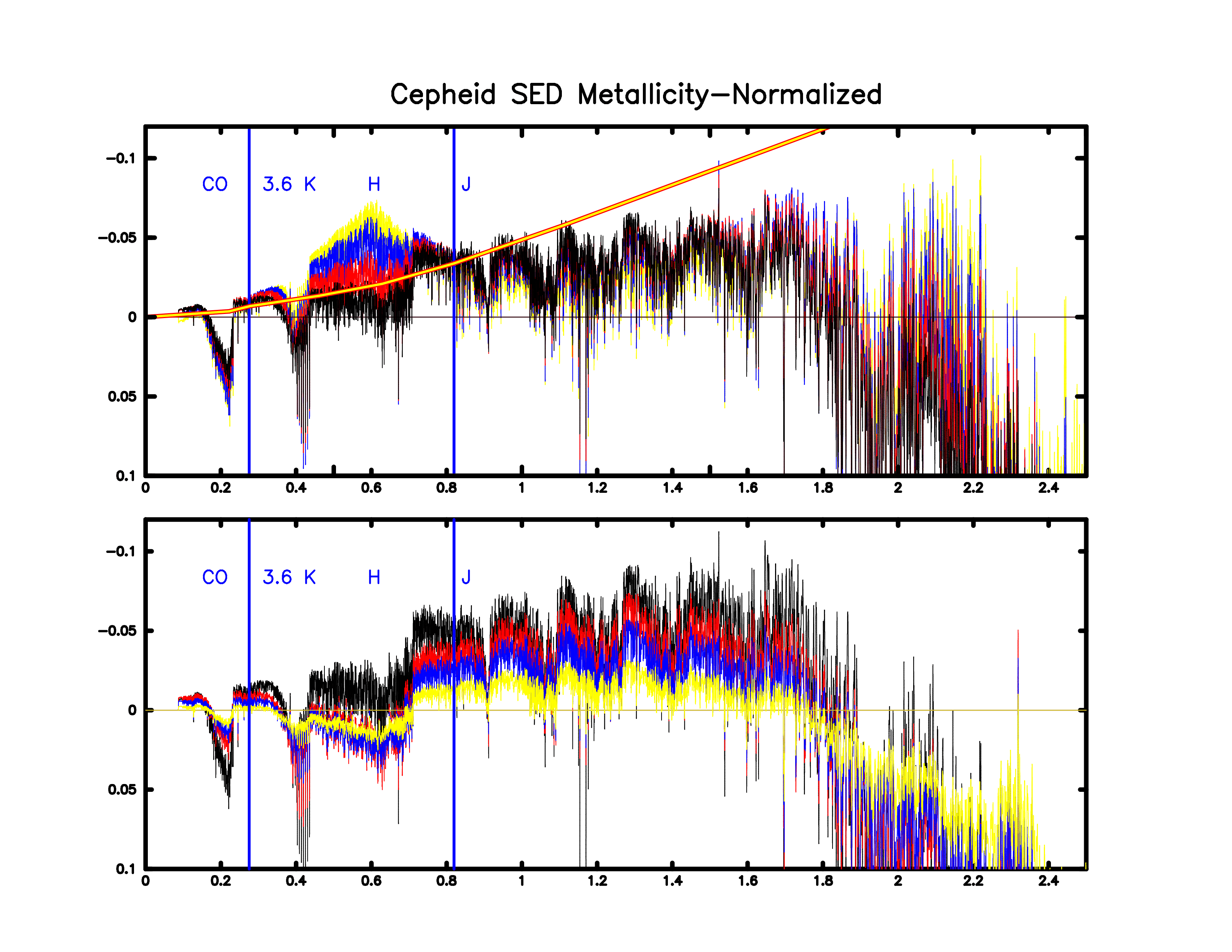}
\caption{\small -- {\bf Lower Panel:}  Four spectra, normalized by a [Fe/H] = +1.0 dex (T = 5000K, log g = 3.5) spectrum as in Figure 3, color-coded by metallicity and over-plotted, each as a function of inverse wavelength. The highest-metallicity spectrum is plotted in black; the lowest-metallicity spectrum is plotted in yellow. {\bf Upper Panel:} Same spectra as found in the lower panel, but scaled and shifted to produce minimum dispersion across the full wavelength range, especially for the two CO molecular bands at 1/4.5 = 0.22 and 1/2.4 = 0.42 inverse microns. The anomalous broadband feature mentioned in Section 3 can be seen centered on the H band, at 1/1.6 = 0.6 inverse microns. The upward trending yellow line is again an interstellar extinction curve fit to the broadband J and 3.6 micron data points, that still extrapolates to zero bias in the true (extinction and metallicity corrected) distance modulus.
        }
\end{figure*}
\begin{figure*}[hbt!]
\centering
\includegraphics[width=18.9cm, angle=-0] {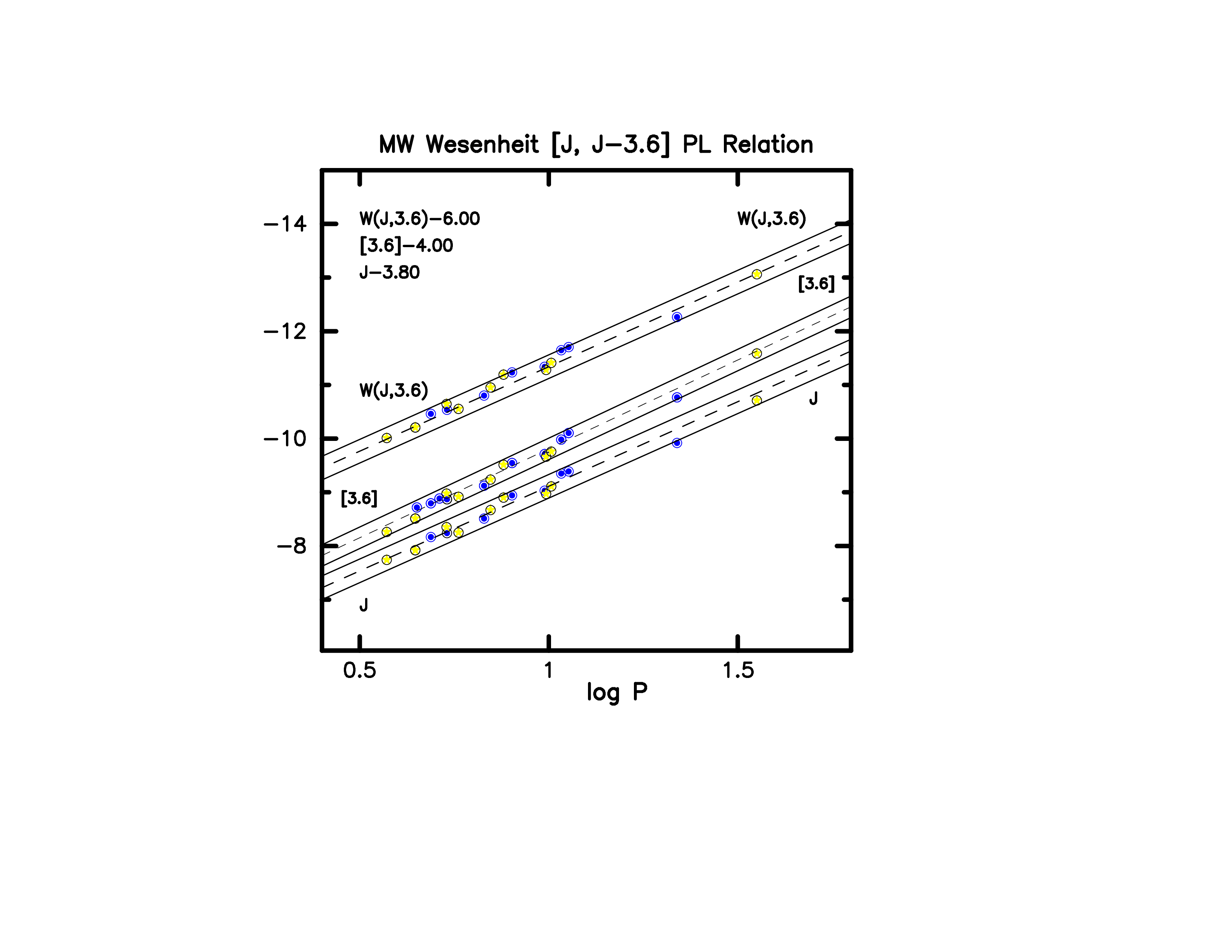}
\caption{\small -- Here we show the J-band and [3.6] micron PL relations in the lower half of the plot, with individual Cepheids color-coded as in Figure 3 of Paper I (Cepheids with HST parallaxes are yellow, those with corrected Gaia parallaxes are blue). Above them is the W(J,3.6) Wesenheit function, which is implicitly corrected both for total line-of-sight interstellar extinction and (stellar-atmospheric) metallicity effects, as discussed in the text. Note how the scatter in W is less than the scatter seen in the J-band PL relation. 
}
\end{figure*}
\clearpage

\begin{figure*}[hbt!] 
\centering
\includegraphics[width=12.9cm, angle=-00] {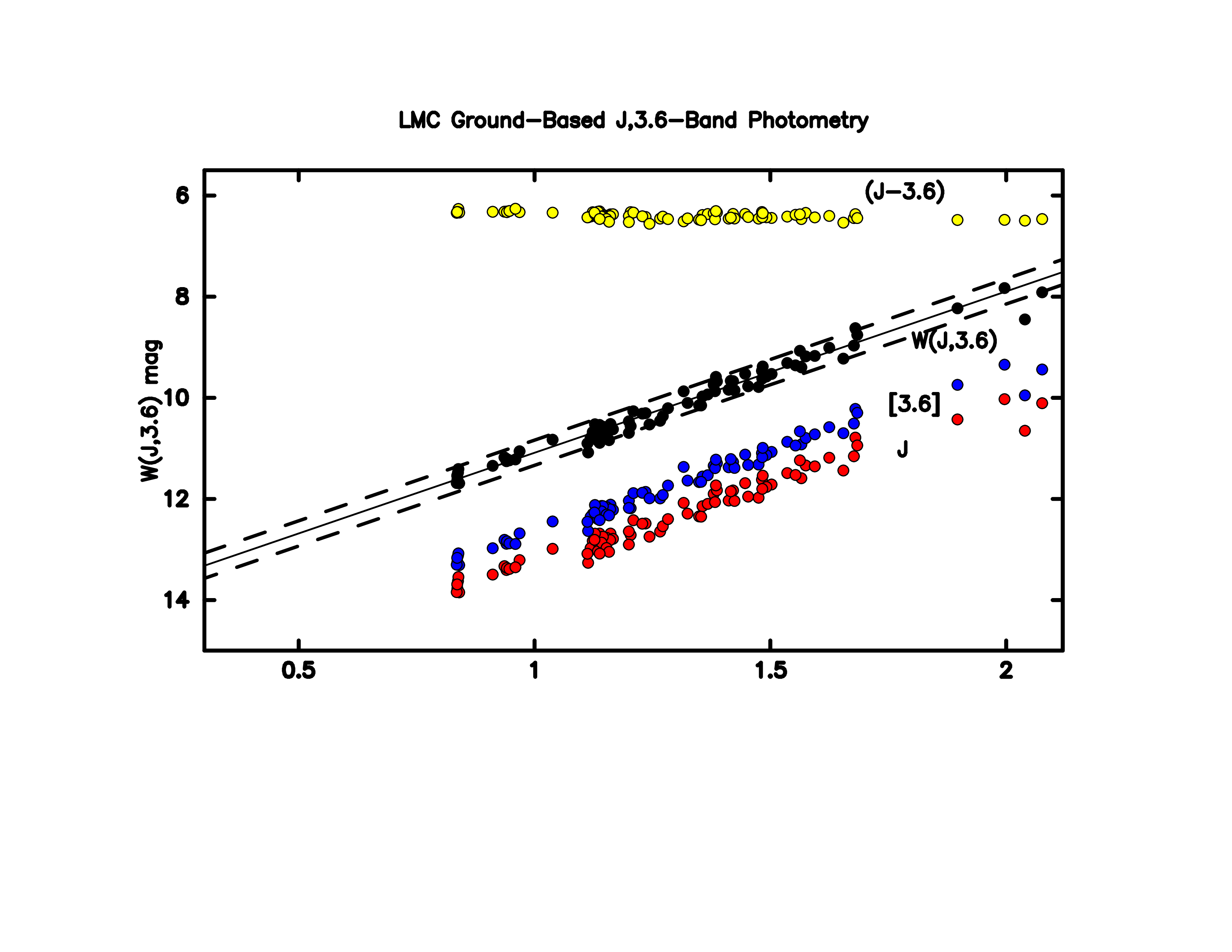}
\includegraphics[width=5.9cm, angle=-90] {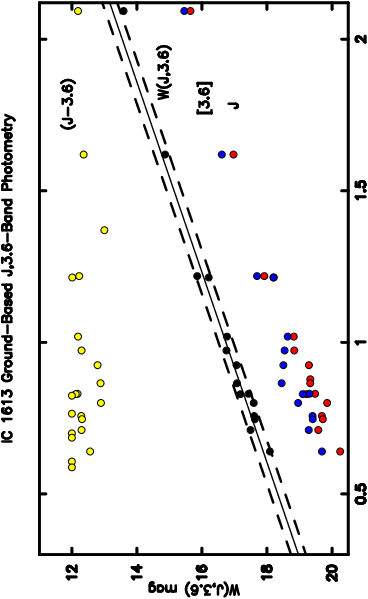}
\caption{\small -- Upper Panel: The reddening-free and metallicity-independent infrared Wesenheit function, W(J,3.6) as a function of period for LMC Cepheids (black filled circles). Blue and red filled circles, respectively show the [3.6] micron and the J-band PL relations used to form W(J,3.6). The extremely tight correlation of the (J-3.6) color as a function of period is shown by the yellow filled circles at the top of the plot. Lower Panel: Same description as for the upper panel, except this plot contains observations of Cepheids in IC~1613. Despite the obviously larger scatter in the (J-3.6) period-color relation, primarily due to lower precision in the [3.6] micron data, the W-logP relation is well behaved in both slope and scatter when compared to the LMC data above. The J and 3.6 PL plots are vertically offset from the W(J,3.6) PL relation
for clarity.}
\end{figure*}

\section{Conclusions}
The discovery of the effect of the CO band head on the broadband [4.5] micron observations of classical Cepheids in the MW, the LMC and the SMC proved, beyond any observational or theoretical doubt, that the atmospheres of these stars do contain measurable signatures of metallicity variations. This is true, despite all of the previous observational evidence gathered to the contrary, that has been discussed in detail in Paper I. The synthetic spectra of Cepheids studied here offer a solution to this apparent paradox: The broadband red (RI) and broad-band near-infrared (JHK) signatures of metallicity are sufficiently degenerate with the effects of interstellar reddening that they are lost in the dereddening procedures currently be employed. That degeneracy is starting to fade at V, quickly increasing in going to the B-band, and further diverging at all shorter wavelengths. 

We have shown that the stellar atmosphere predictions of multi-wavelength metallicity effects can be recovered when an iterative (or simultaneous) fit for reddening and metallicity is undertaken. The past procedure of fitting only for reddening has been hiding the metallicity effects that would otherwise be measurable blueward of the V band.  

In exploring the optimal numbers and combinations of bands required to establish a metallicity and extinction corrected Cepheid distance scale, we have found that a solution can be had with the utilization of two, and only two, NIR and MIR observations of Cepheid.
That combination is a new and very specific form of the original Wesenheit function [~W(V, B-V) = V $-$ R$_V~ \times$ (B-V)~], which was first designed for optical data, being fine-tuned to implicitly/mathematically correct for total line-of-sight interstellar extinction being experienced at those two wavelengths.
The new, long-wavelength Wesenheit function [~W(J, J-3.6) = J $-$
R$_J~  \times$ (J-3.6)~] simultaneously corrects for metallicity and for extinction\footnote{And it also must be mentioned that 
Majaess et al. (2016) found that the ratio of $A_J$ to $A_{3.6}$ was exceptionally well-defined, and stable (as a function of galactic longitude), in contrast to many of the other near- and mid-infrared extinction ratios tested, which ranged from 1.2 to beyond 4.6 microns: their Table 1 and Figures 3 \& 4.}
without any bias in the extrapolated, long-wavelength solution for the true distance modulus, which is, in any case, only a small distance away (in 1/$\lambda$ space) from the defining band passes. That is the good news.

The extremely good news is that both of the above filters are available in the NIRCam imager on board JWST, and they can each be observed in a single exposure, given that the J band can be observed in the NIR detector, while the 4.5 band can be simultaneously observed in the beam-split MIR detector.

\clearpage
\section{Acknowledgements} This research made use of the NASA/IPAC Extragalactic Database (NED), which is operated by the Jet Propulsion Laboratory, California Institute of Technology, under contract with the National Aeronautics and Space Administration. 
We thank the {\it Observatories of the Carnegie Institution for Science} and the {\it University of Chicago} for their past and on-going support of our long-term research into the calibration and determination of the expansion rate of the Universe. Financial support for this work was provided in part by NASA through grant number HST-GO-13691.003-A from the Space Telescope Science Institute, which is operated by AURA, Inc., under NASA contract NAS 5-26555. We thank the anonymous referee for their detailed reading of the manuscript, and for their suggestion that we address the impact on the distance Cepheid distance scale of assuming various values of the ratio of total-to-selective absorption, which we discuss in the Appendix. 

\section{REFERENCES}
\par\noindent
Cardelli, J.A., Clayton, G.C. \& Mathis, J.S. 1989, ApJ, 345, 245 

\par\noindent
Gordon, K.D., Clayton, G.G., Misselt, K.A., et al. 22003, ApJ, 594, 279

\par\noindent
Herrnstein, J.R., Moran, J.M., Greenhill, L.J., et al. Nature, 400, 539

\par\noindent
Husser, T.-O., Wende-von Berg, S., Dreizler, S., et al. 2013, A\&A, 553, 6

\par\noindent
Madore, B.F. 1982, ApJ, 253, 575

\par\noindent
Madore, B.F. \& Freedman, W.L. 2024, ApJ, submitted 2023arXiv230910859M 

\par\noindent
Majaess, D., Turner, D., D\'ek\'any, I., et al., 2016, A\&A, 593, A124

\par\noindent
Pietrzy\'nski, G., Graczyk, D., Gallene, A., et al., 2019, Nature, 567, 200

\par\noindent
Reid, M.J., Pesce, D.W., \& Riess, A.R. 2019, ApJL, 886, 27 

\par\noindent
Turner, D.G. 2014, AAS Poster ``Galactic Extinction, Cepheid Distances, and the Hubble Constant'', 2014AAS...22431809T

\par\noindent
Turner, D.G., Majaess, D.J.\& Balam, D.D. 2014, CaJPh, 92, 1696 

\par\noindent
Zagury, F. 2103, AN, 334, 1107
\vfill\eject

\appendix
\section{The Impact of Changing the Ratio of Total-to-Selective Absorption}
The referee alerted us to papers by Zagury (2013), Turner, Majaess \& Balam (2014), Turner (2014) and many references therein that draw attention to yet another bias in the extragalactic distance scale, whenever line-of-sight extinction with the host galaxy is being corrected for by using multi-band fits to extinction curves, in general, and the use of ``reddening-free" Wesenheit functions, in specific. In either case, the problem confronting the correction for extinction involves the necessity to choose one particular form of the multi-wavelength extinction curve running from the B to V-band optical wavelengths, out into the V and I-band redder wavelengths, and then further into the JHK near infrared, and 3.6 and 4.5 micron mid-infrared bands ** BUT WE DON'T USE THE SAME EXTINCTION CURVE IN THE MID-IR - WE USED INDEBETOUW **. It is well known that interstellar extinction curve (IEC) is not universal, especially in the ultraviolet where a broad 2200 angstrom absorption feature (discovered by Stecher 1965) increases as a function of the host-galaxy metallicity, being strongest in the metal-rich Milky Way IEC and almost absent in metal-poor SMC IEC (see Gordon et al. 2003 for a comprehensive review of the systematic trending of this feature in the extinction curves for the Milky Way, LMC and SMC). To first order, residual effects of metallicity have been explored by changing the optical (B,V band) ratio of the total-to-selective absorption $R_V = A_V/E(B-V)$ over values ranging from 3.2 to 2.8, for high- and low-metallicity galaxies, respectively.

\begin{figure*}[hbt!] 
\centering
\includegraphics[width=8.9cm, angle=-90] {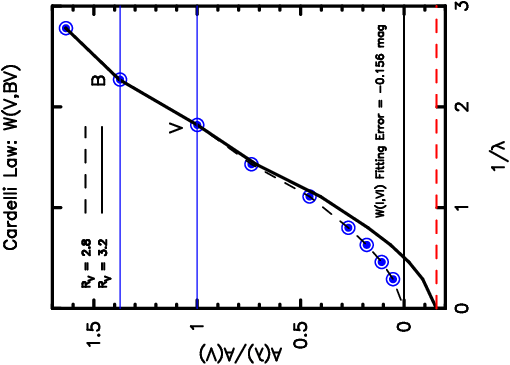}
\caption{\small -- Multi-wavelength apparent distance moduli (circled blue points) for Cepheids having a V-band extinction of $A_V = $ 1.00~mag modulated by an IEC with a ratio of total-to-selected absorption of $R_V = 2.8$. Only the $B$ and $V$ data points are then fit by an IEC with $R_V = 3.1$ (solid black line), which is equivalent to algebraically calculating W(V,BV) adopting $R_V = 3.1$. The difference between the two true moduli is shown by the divergent intercepts at the bottom left of the panel, and offset by -0.156~mag, as emphasized by the horizontal dashed red line.}
\end{figure*}

Turner (2014) explored the impact that ignoring these potential differences between galaxies might have on the extragalactic distance scale, and ultimately the value of the Hubble constant. In his example, BV data are considered where $R_V$ = 3.0 was adopted for the dust in a target galaxy, but R = 3.3 was then used to deredden the data. The dereddening method employed was the original Wesenheit functional form of $W = V - R_V \times (B-V)$ (Madore 1974). Turner's target galaxy was NGC~4258 for which he took the geometric distance modulus of 29.29~mag (7.2 $\pm$0.2~Mpc)  available at that time from Herrnstein et al.(1999),  and compared that to his R = 3.3 solution that gave 29.51~mag, +0.22~mag more distant. This glaring discrepancy led Turner to the conclusion that differences in $R_V$ between the calibrator and the target galaxy could be a dominant (10\%) uncertainty in the extragalactic distance scale.

\begin{figure*}[hbt!] 
\centering\includegraphics[width=8.9cm, angle=-90] {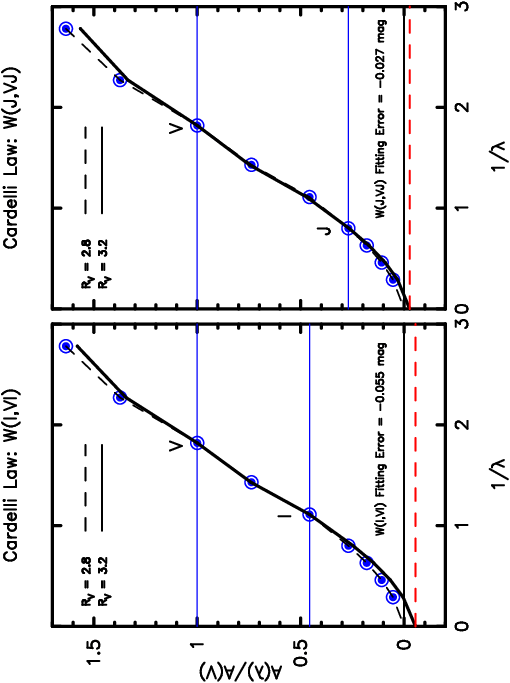}
\caption{\small -- Same as Figure 7 except that the left panel now uses $V$ and $I$ to fit the IEC to those two points giving rise to a reduced offset (dashed horizontal red line at -0.055~mag) between the $R_V =$ 2.8 input and the 3.1 Wesenheit solutions, seen in the lower left corner. The right-hand panel show the further diminished offset (dashed horizontal red line at -0.027~mag) when using W(J,VJ). Horizontal blue line indicate the bands selected for fitting to the $R_V =$ 3.1 IEC.}
\end{figure*}

In the intervening 10 years much has changed to ameliorate that concern. First, the geometric distance modulus to NGC~4258 has been increased to 29.479~mag (Reid et al. 2021) cutting Turner's (2014) discrepancy in half, down to an offset of ``only" 0.11~mag. Second, concerning the search for and discovery of new Cepheids for the derivation of PL relations for nearby galaxies from the HST Key Project onward, no major surveys have acquired or used B-band data in the correction for line-of-sight extinction.  

But the question still remains: Is the value of $R_V$, and its potential variations from galaxy to galaxy, still a major source of statistical and/or systematic uncertainty in the extragalactic distance scale? To address that question we first start with Figure 7, which is an expanded, multi-wavelength view of Turner's (2014)  investigation into the changes in true moduli due to incorrect adoption of extinction curves applied to $BV$ apparent moduli, and parameterized by $R_V$. The circled blue data points show nine multi-wavelength extinctions (U in the upper right corner, 3.6~microns in the lower left corner) each of which is scaled to a V-band extinction of 1.00~mag, adopting a host-galaxy value of $R_V (V,BV) =$ 2.8. The dashed line, interpolating each of the discrete points is the generating interstellar extinction curve, using the prescription given by Cardelli, Clayton \& Mathis (1989). The solid black line is the IEC scaled to  $R_V = $ 3.1 and fit only to the two B and V data points (emphasized by thin blue horizontal lines). That IEC fit to only two points is mathematically equivalent to the Wesenheit solution, $W(V,BV) = V - 3.2 \times (B-V)$.
That $R_V = 3.2$ fit results in a calculated true distance modulus offset by -0.156 mag from the $R_V = 2.8$ solution. This is simply a restatement and general confirmation of Turner's original assertion. Nevertheless, it is worth noting that this is an extreme case in which the calibrator is at one end of the IEC range, and the target galaxies are all at the other extreme; and the effect would be systematic in nature. Alternatively, should the calibrator be in the middle of the range and the target galaxies scattering above and below that value, then the effect would be reduced to being statistical in nature.

The left panel in Figure 8 now sheds some quantitative light on that same question of IEC variation, but now for the W(V,VI) solution used in the HST Hubble Constant Key Project (Freedman et al. 2001).  The fit is forced through the V and I moduli as again highlighted by the two horizontal blue lines. The difference between the two $R_V$ solutions has now dropped by a nearly factor of 3, when compared to the W(V,BV) solution: down from -0.156~mag to -0.055~mag. However, it is still larger than the 1\% extragalactic distance-scale goal of many Hubble constant programs using Cepheids.  

The right panel of Figure 8 then offers a two-band $W(J,JI)$ solution, introducing JWST/NIRCam data. That solution reduces the impact of plausible variation in the value of $R$ on the true distance modulus by another factor of two; down to a bias of only -0.027~mag. With this combination we are now at the 1.4\% level for the maximum impact of the incorrect assumption of the value of the ratio of total-to-selective absorption, when used to deredden Cepheids employing near-infrared Wesenheit magnitudes. 
\vfill\eject
However, if more than two bands are available to be fit by a multi-wavelength extinction curve, we then suggest the following: (a) First, individually correct each of the bands for the independently published foreground extinctions for that galaxy as a whole, using the most up-to-date, Milky Way value of $R_V$. (b) Plot the new apparent distance moduli, and then select the two parameters, $A_I$ and $R_V$  that, in combination, minimize the scatter of those apparent moduli around a selected extinction curve. As can be seen in the right panel of Figure 8 especially, the deviations in any given band will be small, but they will be systematic (as more readily seen as a function of wavelength in the left panel of Figure 8). In addition, they will almost certainly require higher-precision data than is currently available in the visual photometry. If 1\% distances are required to support a 1\% value of the Hubble constant, then higher precision data will be essential in producing a systematically unbiased cosmological solution. 
\begin{figure*}[hbt!] 
\centering\includegraphics[width=8.9cm, angle=-90] {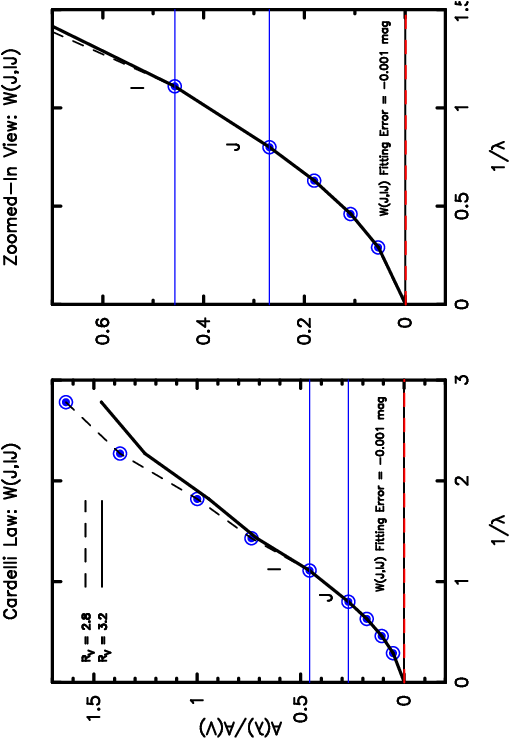}
\caption{\small -- Same as Figure 7 except that the left panel uses $J$ and $I$ to fit the IEC to those two points giving a vanishingly small offset (of -0.001~mag) between the $R_V =$ 2.8 input and the 3.1 Wesenheit solution, barely visible in the lower left corner. The right-hand panel shows a zoomed-in of the region around the intercept in left-hand panel. Clearly, fitting either of the extinction curves to any to data points from I to 3.6 microns will not be biased at any statistically significant. JWST/NIRCam can access any such pairs of magnitudes without recourse to HST observations.}
\end{figure*}

In the absence of multi-color apparent moduli, or in the interest of establishing a purely JWST-based distance scale, Figure 9 shows what only two JWST/NIRCam bands can deliver. This far into the infrared, the $W(J,JI)$ solution is indistinguishible from the full SED fit, regardless of which value of $R_V$ is chosen to deredden the moduli. The formal offset is only -0.001~mag level.
Thus, JWST data can  solve this problem.

\end{document}